\documentclass[journal]{IEEEtran}

\usepackage{graphicx}
\usepackage{epsf}
\usepackage{mathtools}
\usepackage{amssymb}
\usepackage{amsmath, amsfonts}
\usepackage{latexsym}
\usepackage{multirow}
\usepackage{multicol}
\usepackage[table]{xcolor}
\usepackage{color}
\usepackage{booktabs}

\usepackage{tabularx}
\usepackage{lipsum}

\usepackage{algorithm}
\usepackage{algorithmic}

\usepackage{mathrsfs}

\usepackage{fixltx2e}

\usepackage[mathscr]{euscript}

\usepackage{commath}
\usepackage{MnSymbol}

\usepackage{subcaption}

\usepackage{mathtools} 
\DeclarePairedDelimiterX\setc[2]{[}{]}{\,#1 \;\delimsize\vert\; #2\,}
\DeclarePairedDelimiterX\parth[2]{(}{)}{\,#1 \;\delimsize\vert\; #2\,}

\PassOptionsToPackage{hyphens}{url}
\usepackage[unicode=true, bookmarks=false, bookmarksnumbered=true, bookmarksopen=true, bookmarksopenlevel=1, breaklinks=false, pdfborder={0 0 0}, pdfborderstyle={}, backref=false, colorlinks=false]{hyperref}

\usepackage{theorem}
{
\theorembodyfont{\rmfamily}

}

\usepackage{soul}

\definecolor{orange}{RGB}{255,127,0}
\definecolor{blue}{RGB}{0,0,255}
\definecolor{red}{RGB}{255,0,0}
\definecolor{green}{RGB}{50,160,50}
\definecolor{grey}{RGB}{125,120,125}
\definecolor{purple}{RGB}{125,0,125}

\begin{document}
{
\title{{\fontsize{16}{2}\selectfont Smart Contract Tells: Aircraft Maintenance Records Are Now Trustworthy}}

\author
{
Woosuk Choi and Seungmo Kim, \textit{Senior Member}, \textit{IEEE}

\vspace{-0.3 in}

\thanks{W. Choi is with Chung-Ang University, Seoul, South Korea, who can be reached at oosukchoi@cau.ac.kr. S. Kim is with Georgia Southern University, Statesboro, GA, USA. The corresponding author of this paper is S. Kim who can be reached at seungmokim@georgiasouthern.edu. This work was supported by Georgia Southern University.}
}

\maketitle

\begin{abstract}
Aircraft maintenance records are critical to airworthiness and asset valuation, yet they are often fragmented across stakeholders, creating verification bottlenecks and information asymmetry that may suppress aircraft residual value. This paper proposes a blockchain-anchored decentralized application (dApp) based on a dual-layer architecture that combines InterPlanetary File System (IPFS)-based off-chain storage with on-chain anchoring of Content Identifiers (CIDs) and selected metadata. With respect to off-chain file size, the on-chain payload per record remains $\mathcal{O}(1)$, compared with $\mathcal{O}(n)$
for direct on-chain file storage. The architecture incorporates metadata and traceability controls informed by Federal Aviation Administration (FAA) electronic recordkeeping guidance. The main contribution is an economic framework that models the relationship between tamper-evident maintenance-record provenance, audit workflow duration, aircraft residual value, and operational cost. In a 7-kB experiment conducted on the BNB Smart Chain testnet, CID anchoring reduced gas consumption by 93.9\% compared with direct on-chain storage. Under explicitly stated scenario assumptions, the audit-cost model indicates potential savings of more than 90\%. These results support the technical feasibility of the prototype and illustrate its economic potential, while the estimated financial benefits remain to be validated using operational data.
\end{abstract}


\section{Introduction}
\subsection{Background}
Aircraft maintenance records are more than just logs; they are critically fundamental to airworthiness, regulatory compliance, and long-term valuation of an aviation asset. Although the United States Federal Aviation Administration (FAA) Advisory Circular (AC) 120-78B provides a framework for electronic recordkeeping to ensure authenticity and traceability, the reality on the ground often gets complicated \cite{ac120}.

In practice, maintenance data are scattered across a fragmented landscape of operators, maintenance, repair, and overhaul (MRO), and lessors. This isolation leads to a variety of problems including data format inconsistency and interoperability obstacles, leading to inefficiency in subsequent verifications \cite{iata2022}. This fragmentation is most felt during high-stake periods such as lease transitions or audits, where back-to-birth (BtB) traceability is quintessential \cite{iata2025}. However, the industry is still relying on manual cross-referencing of records across relevant systems, which is often prone to human error and operational bottlenecks \cite{iata2022} and thus leads to opportunity costs on owners and operators \cite{iba2024}.
 
As a solution to this inefficiency, we introduce a blockchain-based management system that provides a ``shared tamper-evident integrity reference'' for maintenance data. The proposed system features a dual-layer architecture, which offers a way to anchor records in an immutable manner with a significantly reduced overhead of blockchain storage.

\subsection{Related Work}
Earlier literature of aircraft maintenance records mainly addressed physical logbooks to identify vulnerabilities such as susceptibility to loss, destruction, and signature forgery \cite{aleshi2020}. Distributed ledgers emerged as a solution to such inefficiency. One example is AirChain, a blockchain-based system that stores maintenance records in a tamper-resistant and readily accessible manner, with experimental validation on a live network \cite{jensen2022}. The literature was extended to integrating the InterPlanetary File System (IPFS) with blockchain to simultaneously address traceability, data integrity, and privacy for aviation records \cite{ipfsaviation2023}. All in all, this prior work establishes the foundational architecture for decentralized maintenance recordkeeping but stop short of modeling the financial implications of verification latency.
 
Meanwhile, smart contract has been identified as a solution establishing trust in other applications. A blockchain-based electronic health record systems was proposed as a design reference for multi-stakeholder, non-repudiation environments \cite{rahman2022}. In supply chain literature, smart contracts emerged as a method promoting traceability with adaptive gas fee mechanisms that maintain scalability under high transaction volumes \cite{smartcontractsc2024}. Reducing on-chain storage overhead, dual-blockchain IPFS framework has demonstrated the complexity of $\mathcal{O}(1)$ satisfying preserved cryptographic verifiability \cite{zareen2025}. Moreover, decentralized application (dApp) caching architectures have confirmed performance and integrity benefits in multi-stakeholder deployments \cite{kim2024}.
 
Recently, the literature of aviation embodied integration of blockchain, with argument that record quality directly affects airworthiness decisions. A recent proposal adopted Hyperledger Fabric for traceability of spare parts \cite{ho2021}. A dApp was proposed for promotion of immutability across manufacturers, vendors, MROs, and airlines \cite{richardo2022}. A latest study quantitatively analyzed the economic benefits of distributed ledger adoption in aviation materials management \cite{zhang2024}. One can find an even broader study on integration of artificial intelligence (AI) into blockchain for MRO records management \cite{aiblockchain2022}, which was lately specified to pairing on-chain anchoring with AI-driven visual inspection for engine blade traceability \cite{bladechain2025}.
 
Even beyond maintenance, blockchain has been adopted in other key areas of aviation as the main method of promoting data integrity. Examples include applications of blockchain to unmanned aircraft system traffic management (UTM) security for unmanned aircraft systems \cite{utmsafety2024}, layer-2 auditability frameworks for autonomous multi-unmanned aerial vehicle (UAV) operations \cite{uavteamwork2023}, and performance characterization of permissioned ledgers for advanced air mobility \cite{nunes2024}.

\subsection{Contribution}
Prior studies have established the technical feasibility of using blockchain and IPFS to improve the integrity and traceability of aircraft maintenance records \cite{jensen2022,ipfsaviation2023,zareen2025}. Accordingly, this work does not claim novelty in the underlying blockchain, smart-contract, or distributed-storage primitives. Instead, its primary contribution is an application-oriented economic framework that examines how tamper-evident record provenance and reduced record-verification effort may translate into financial outcomes in aircraft operations and asset management. The implemented decentralized application serves as a proof-of-concept platform for operationalizing and evaluating this framework. The main contributions are summarized as follows:

\begin{itemize}
\item It formulates a residual-value sensitivity model that captures the financial effects of maintenance-record completeness, provenance, and verification delay. The model focuses on the financial recognition of documented maintenance value rather than treating blockchain validation as proof that the underlying maintenance activity was physically performed.

\item It develops a scenario-based audit-cost model that relates record retrieval, integrity verification, and Back-to-Birth (BtB) traceability workloads to labor cost, hangar occupancy, aircraft downtime, and the associated opportunity cost.

\item It implements and evaluates a blockchain--IPFS dApp that supports maintenance-file anchoring, retrieval, integrity checking, transaction traceability, and residual-value visualization. The architecture stores maintenance files off-chain and anchors fixed-length CIDs and selected metadata on-chain, resulting in $\mathcal{O}(1)$ per-record on-chain storage complexity with respect to the underlying file size. In a 7-kB experiment conducted on the BNB Smart Chain testnet, CID anchoring reduced gas consumption by 93.9\% relative to direct on-chain storage.
\end{itemize}

The economic results are presented as scenario-based estimates of potential benefits rather than field-validated financial outcomes.

\section{Proposed System Architecture: Dual-Layer Trust Model}

\subsection{Architecture Overview}

The proposed system is implemented as a dApp designed to manage aircraft maintenance records under a blockchain-anchored integrity framework.
The architecture follows a dual-layer trust model, separating high-volume technical data storage from immutable integrity anchoring. 

The system consists of four principal components:

\begin{enumerate}
    \item Data Input and Standardization Layer: Standardized maintenance datasets (CSV format) aligned with FAA AC 120-78B \cite{ac120} and ATA Spec 2000 \cite{spec2000} are uploaded by authorized stakeholders (e.g., MRO facilities, operators). Each dataset contains structured metadata including Aircraft Registration, Time in Service (TIS), Flight Cycles (FC), maintenance event type, and authorized technician signature in accordance with 14 CFR Part 43.

    \item Off-Chain Storage Layer (IPFS using Pinata): To avoid the prohibitive costs of storing large maintenance files directly on the blockchain, the system utilizes the IPFS, a content-addressed distributed storage network \cite{zareen2025}. Once a raw maintenance file is stored in the IPFS, it generates a unique Content Identifier (CID):
    \begin{equation} 
    CID = H(F)
    \end{equation}
    where $H(\cdot)$ represents a cryptographic hash function (e.g., SHA-256) and $F$ is the maintenance file. Any modification to the file produces a different CID, ensuring intrinsic data integrity. To guarantee data persistence and availability, a pinning service (e.g., Pinata) is used to maintain redundancy across distributed nodes.
    \item On-Chain Smart Contract Layer: The system anchors a concise metadata schema consisting of the aircraft tail number, model type, initial valuation, and mechanic license credentials (per FAA 14 CFR Part 65). Crucially, it stores the Content Identifiers (CIDs) for both the raw and processed datasets, providing a tamper-proof cryptographic link to the off-chain storage. This creates a permanent and immutable anchor that links the off-chain dataset to a blockchain-verified event.
    \item Retrieval and Audit Interface: Authorized stakeholders query the smart contract using a unique File ID. The system retrieves the corresponding CID, fetches the file from IPFS, and recomputes the hash to verify integrity before reconstruction of the maintenance history.
\end{enumerate}

This separation establishes a minimal on-chain footprint while preserving cryptographic verifiability. Figure 1 visualizes the process flow.

\begin{figure}[t]
\centerline{\includegraphics[width=\columnwidth]{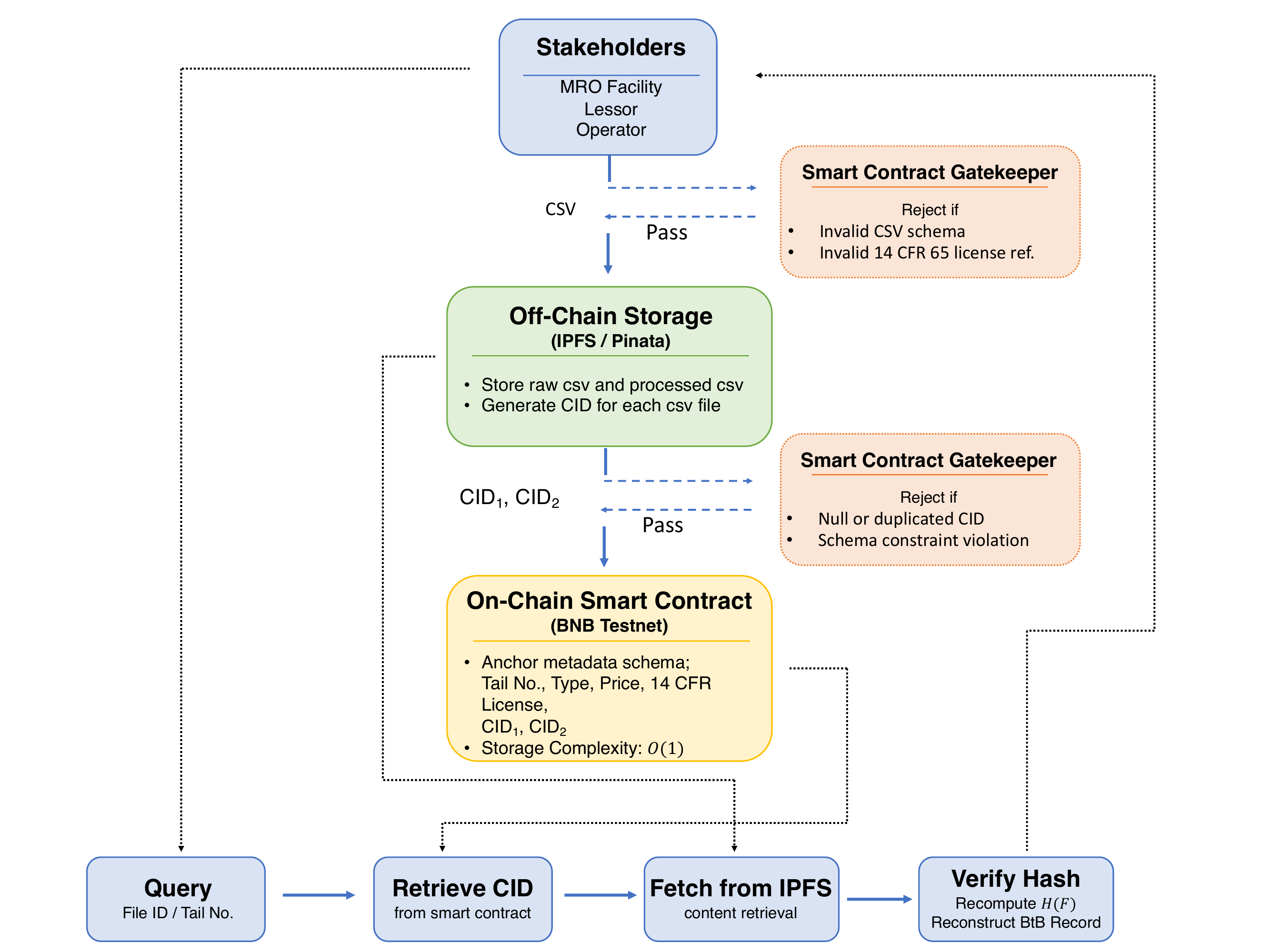}}
\caption{Proposed system's 4-layer architecture}
\label{fig1}
\end{figure}

\subsection{Storage and Gas Complexity}
A critical constraint in blockchain-based systems is storage cost. On Ethereum-compatible networks, storing raw data directly on-chain incurs gas costs proportional to the data size. If $n$ denotes the file size in bytes, the complexity for complete on-chain storage is $O(n)$, while the CID-based storage strategy reduces complexity to $O(1)$ by anchoring only a fixed-length cryptographic hash on-chain. Figure 2 illustrates this theoretical complexity divergence alongside empirically measured gas consumption values, confirming that the CID-anchoring approach maintains constant cost regardless of file size.

In experimental validation on the BNB Testnet, storing a 7 KB maintenance CSV directly on-chain required 7,284,350 gas consumption. On the other hand, recording only metadata schema and two CIDs required 443,301 gas consumption. This corresponds to a 93.9\% reduction in transaction cost. By fixing the on-chain footprint to a constant-length string, the system enables scalability across fleet-level operations where annual maintenance logs may exceed gigabytes of structured records.

Figure 3 presents a screenshot of the implemented dApp following the successful anchoring of a nine-year aircraft maintenance record, demonstrating the system's retrieval interface, real-time residual value visualization, and on-chain transaction ledger.

\begin{figure}[t]
\centerline{\includegraphics[width=\columnwidth, height = 6.5cm]{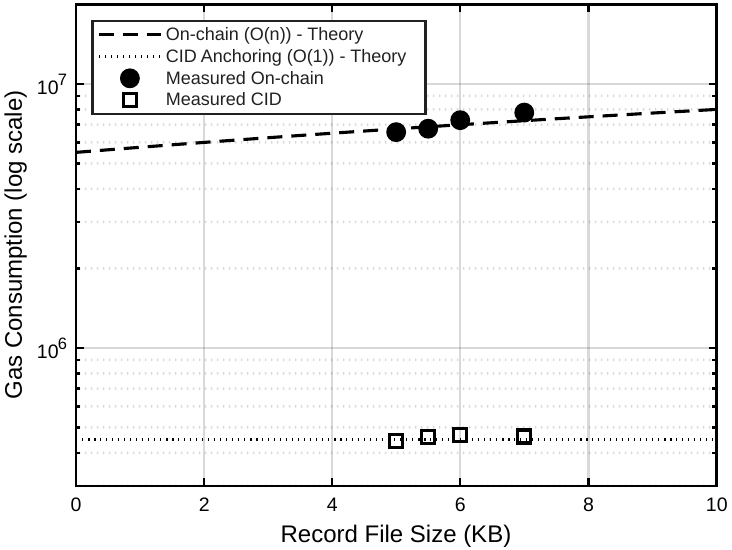}}
\caption{Gas consumption complexity comparison}
\label{fig2}
\end{figure}

\begin{figure}[t]
\centerline{\includegraphics[width = \columnwidth, height = 7cm]{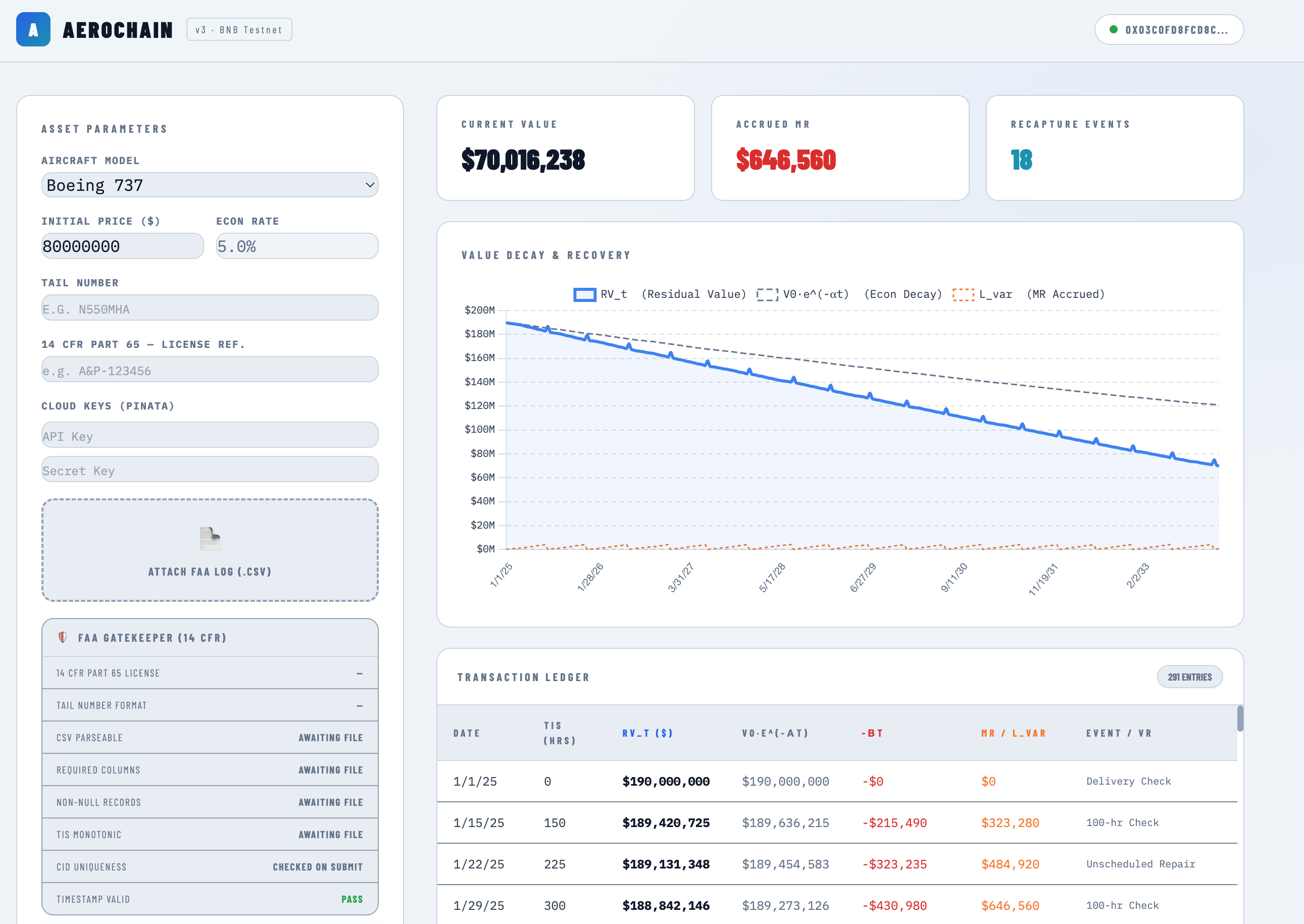}}
\caption{Screenshot of proposed dApp after anchoring a 9-year operated maintenance record sample}
\label{fig3}
\end{figure}

\subsection{Compliance-Oriented Metadata and Traceability Controls}
The architecture is designed to align selected metadata and traceability controls with the electronic-recordkeeping guidance of FAA AC 120-78B., which mandates that electronic recordkeeping systems provide a high degree of confidence in authenticity and non-repudiation. Maintenance datasets include metadata fields informed by 14 CFR Part 43, such as Aircraft Registration, Airframe Total Time in Service (TIS), Flight Cycles (FC), and a technician signature reference. By anchoring the hash of each maintenance dataset on-chain, the system creates a timestamped and immutable compliance reference consistent with regulatory traceability requirements.

The smart contract functions as an automated gatekeeper prior to record anchoring. Submissions are programmatically rejected if they fail to satisfy predefined schema constraints, including a registered 14 CFR Part 65 mechanic-license reference, a non-null cryptographic Content Identifier (CID), a unique File ID mapping, and timestamp verification. This validation mechanism reduces human-induced data entry errors and prevents anchoring by unregistered submitters from being permanently embedded within the Back-to-Birth (BtB) record structure.

While blockchain guarantees immutability after record confirmation, it does not resolve the oracle problem at the data-input layer. The
smart contract validates only submitter authorization and predefined metadata constraints; it cannot observe whether a maintenance task
was physically performed or whether the reported values are factually correct. Consequently, an authorized but malicious or compromised
submitter may anchor a well-formed but false maintenance record. The proposed system therefore provides post-submission tamper evidence,
timestamping, and provenance, but not ground-truth certification of the underlying maintenance event. An on-chain record must not, by
itself, be interpreted as sufficient evidence for a residual-value adjustment. Accordingly, integrity assurance is expressed as:

    \begin{equation} 
    P_{tamper} \le P_{hash} + P_{key} + P_{input}
    \end{equation}
where $P_{hash}$ represents the probability of cryptographic hash collision, $P_{key}$ represents the probability of private key compromise, and  $P_{input}$ represents the probability of malicious or erroneous data submission. This formulation provides a conservative risk threshold, ensuring system security even under potential dependencies between threat vectors.

Under SHA-256 assumptions, the collision probability is bounded by:
    \begin{equation} 
    P_{hash} \approx 2^{-256}
    \end{equation}
which is computationally negligible in practical aviation contexts. Unlike $P_{\mathrm{hash}}$, the input-risk term $P_{\mathrm{input}}$ is not reduced by cryptographic hashing or blockchain consensus and is not empirically quantified in this prototype. Therefore, the negligible SHA-256 collision probability should not be interpreted as a negligible probability that the underlying maintenance information is factually incorrect. Mitigating this residual risk requires external controls, such as independent MRO endorsement, trusted mechanic-credential registries, multi-party approval, or sensor-based attestations, which are outside the scope of the current implementation. Consequently, the proposed dual-layer architecture makes post-submission modification detectable under the stated cryptographic assumptions, while source-data validity and private-key compromise remain external operational risks.

\section{Dynamic Residual Value Quantification Model}
\subsection{Framework Motivation and Value Recapture Logic}
Aircraft residual value (RV) estimation has traditionally been derived from deterministic depreciation schedules combined with periodic technical inspections. Industry-standard appraisal frameworks, such as those referenced by IATA and Eurocontrol \cite{iata2016}, \cite{eurocontrol2020}, model aircraft value decay as a function of economic depreciation, structural fatigue, and component wear. These models implicitly assume that maintenance documentation is complete, authentic, and verifiable at the time of valuation.

However, in real-world aircraft transactions - particularly during lease termination and redelivery - the valuation process is not purely mechanical. It is influenced by information asymmetry regarding the completeness and credibility of maintenance records. When Back-to-Birth (BtB) traceability is fragmented across multiple data silos or delayed due to manual verification, market participants incorporate an uncertainty premium into the pricing process. As a result, residual value becomes partially contingent upon record transparency rather than solely on physical aircraft condition.

To formalize this concept, aircraft residual value at time $t$ can be expressed as a function of three interacting dimensions:
    \begin{equation} 
    RV_t = f(D_{economic}, D_{physical}, T_{transparency})
    \end{equation}
where $D_{economic}$ represents time-based economic depreciation, $D_{physical}$  represents usage-driven physical degradation, and $T_{transparency}$ represents the degree of verifiable maintenance integrity.

Conventional appraisal frameworks model only the first two components explicitly. The third dimension—transparency—is typically embedded implicitly within negotiated discounts or risk spreads during secondary market transactions.
The proposed blockchain-anchored maintenance architecture transforms transparency from an implicit assumption into a quantifiable parameter. As established in Section II, the probability of undetected record compromise is bounded by cryptographic and operational factors. By reducing integrity uncertainty through smart contract validation and immutable anchoring, the system reduces verification latency and narrows the information asymmetry gap.

Within this framework, aircraft value degradation can be decomposed into two categories; permanent depreciation, and  variable (restorable) depreciation. Permanent depreciation is irreversible structural fatigue and long-term economic aging, and variable depreciation wears on life-limited parts (LLPs) and components that can be restored through scheduled maintenance events.

The concept of Value Recapture is defined as the financial recognition of previously accumulated variable depreciation after the maintenance event has been accepted through appropriate off-chain procedures and the integrity of its supporting record has been verified. Formally:
    \begin{align}
        VR_t &=\delta\cdot L_{variable,t}
    \end{align} 
    
where, $L_{variable,t}$ represents accumulated restorable loss, and $\delta \in [0,1]$ represents the restoration coefficient determined by maintenance scope and verification completeness.

In conventional recordkeeping systems, $\delta$ may be effectively less than 1 during the verification window due to documentation uncertainty. In the present framework, $\delta$ approaches unity after the underlying maintenance event and its supporting record have been accepted through appropriate off-chain technical and organizational procedures. Blockchain anchoring may reduce the subsequent verification delay, but successful smart-contract validation alone does not establish the physical completion or factual correctness of the maintenance activity. Importantly, this mechanism does not artificially increase intrinsic aircraft value. Instead, it aligns financial valuation timing with physical maintenance reality by removing verification latency. The framework therefore shifts residual value modeling from a purely depreciation-based structure toward an integrity-adjusted dynamic valuation model. This establishes the theoretical foundation for the deterministic mathematical formulation presented in following section and the stochastic uncertainty modeling.

\subsection{Mathematical Formulation}
To formalize the proposed dynamic valuation framework, the aircraft residual value (RV) is modeled as the net result of economic depreciation, permanent physical loss, and restorable variable loss.

Let $V_0$ denote the initial aircraft market value at delivery, $t$ denote Time in Service (TIS), and $VR_t$ denote the value recapture at time $t$.

The residual value at time $t$ is expressed as:
    \begin{align}
    Rv_t &= V_0e^{-\alpha t} - \beta t - \gamma t + VR_t
    \end{align}
where, 
\begin{align*}
    VR_t &=\delta\cdot L_{variable,t} \\ 
    L_{variable,t} &= \gamma t
\end{align*}

Thus, 
    \begin{align}
        Rv_t &=  V_0e^{-\alpha t} - \beta t - (1-\delta)\gamma t
    \end{align}
    
where $\alpha$ is continuous economic depreciation rate, $\beta$ is permanent physical depreciation per unit TIS, $\delta$ is restoration coefficient, and $\gamma$ is variable physical depreciation per unit TIS. 

The first term, $V_0 e^{-\alpha t}$, represents market-driven aging effects and macroeconomic valuation decline. The exponential form reflects proportional decay consistent with financial asset pricing theory and is analytically convenient for stochastic extension.  Let a is annual economic depreciation rate, then $\alpha$ is formulated by:
    \begin{align*}
    \alpha &= -ln(1-r)
    \end{align*} 
where $r$ is the annual depreciation rate.

When the annual depreciation rate $r$ is 5.0\%, the constant $\alpha$ is approximately 0.051. The second term, $\beta t$,  represents irreversible structural fatigue and long-term aging that cannot be restored through maintenance intervention. $\beta$ is calculated 40\% of maintenance reserve (MR). The third term, $(1-\delta)\gamma t$, captures the valuation impact of variable wear that is restorable but not yet financially recognized. $\gamma$ is calculated by 60\% of maintenance reserve (MR). For independently accepted maintenance records, blockchain anchoring may reduce subsequent verification latency and documentation-related valuation suppression. 

To enable quantitative evaluation, the model parameters are calibrated based on industry-standard aircraft valuation and cost data. Specifically, the initial value, economic depreciation rate, operating cost, and maintenance reserve parameters are derived from established aviation financial references, including IATA and EUROCONTROL guidelines \cite{iata2016}, \cite{eurocontrol2020}.  The resulting parameter values for representative Boeing and Airbus aircraft are summarized in Table I and Table II.

\begin{table}[htbp]
\centering
\caption{Constant Values for Boeing Aircraft}
\label{tab:boeing_constants}
\begin{tabular}{l|cccc}
\toprule
 &\textbf{B737} & \textbf{B767} & \textbf{B777} & \textbf{B787} \\ \midrule
Initial value & 80M & 150M & 140M & 190M \\
Operating cost / TIS  & 4337 & 6675 & 9507 & 7184 \\
Maintenance Reserve & 2168.5 & 3337.5 & 4753.5 & 3592 \\
$\alpha$ & 0.051 & 0.051 & 0.051 & 0.051 \\
$\beta$ & 867.4 & 1335 & 1901.4 & 1436.8 \\
$\gamma$ & 1301.1 & 2002.5 & 2852.1 & 2155.2 \\ \bottomrule
\end{tabular}
\end{table}

\begin{table}[htbp]
\centering
\caption{Constant Values for Airbus Aircraft}
\begin{tabular}{l|cc}
\toprule
 &\textbf{A320} & \textbf{A330} \\ \midrule
Initial value [USD] & 77M & 175M \\
Operating cost / TIS & 4829 & 7827 \\
Maintenance Reserve & 2414.5 & 3913.5 \\
$\alpha$ & 0.051 & 0.051 \\
$\beta$ & 965.8 & 1565.4 \\
$\gamma$ & 1448.7 & 2348.1 \\ \bottomrule
\end{tabular}
\end{table}

\subsection{Sensitivity Analysis}

\begin{figure}[t]
\centerline{\includegraphics[width = \columnwidth, height = 18cm]{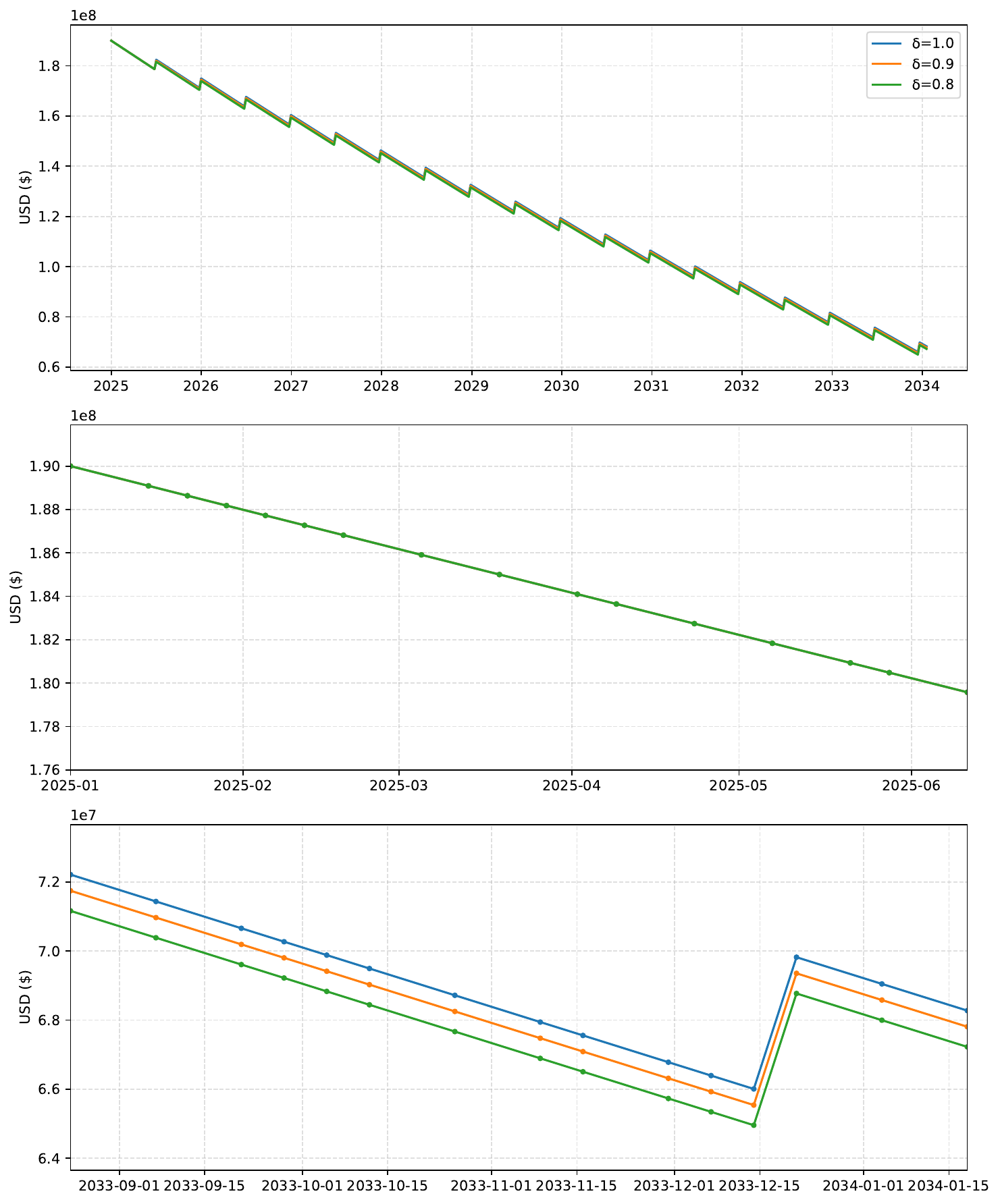}}
\caption{Residual value sensitivity analysis regarding record transparency ($\delta$) with zoomed-in views of the initial and final phases.}
\label{fig4}
\end{figure}

\begin{figure}[t]
\centerline{\includegraphics[width = \columnwidth, height = 6cm]{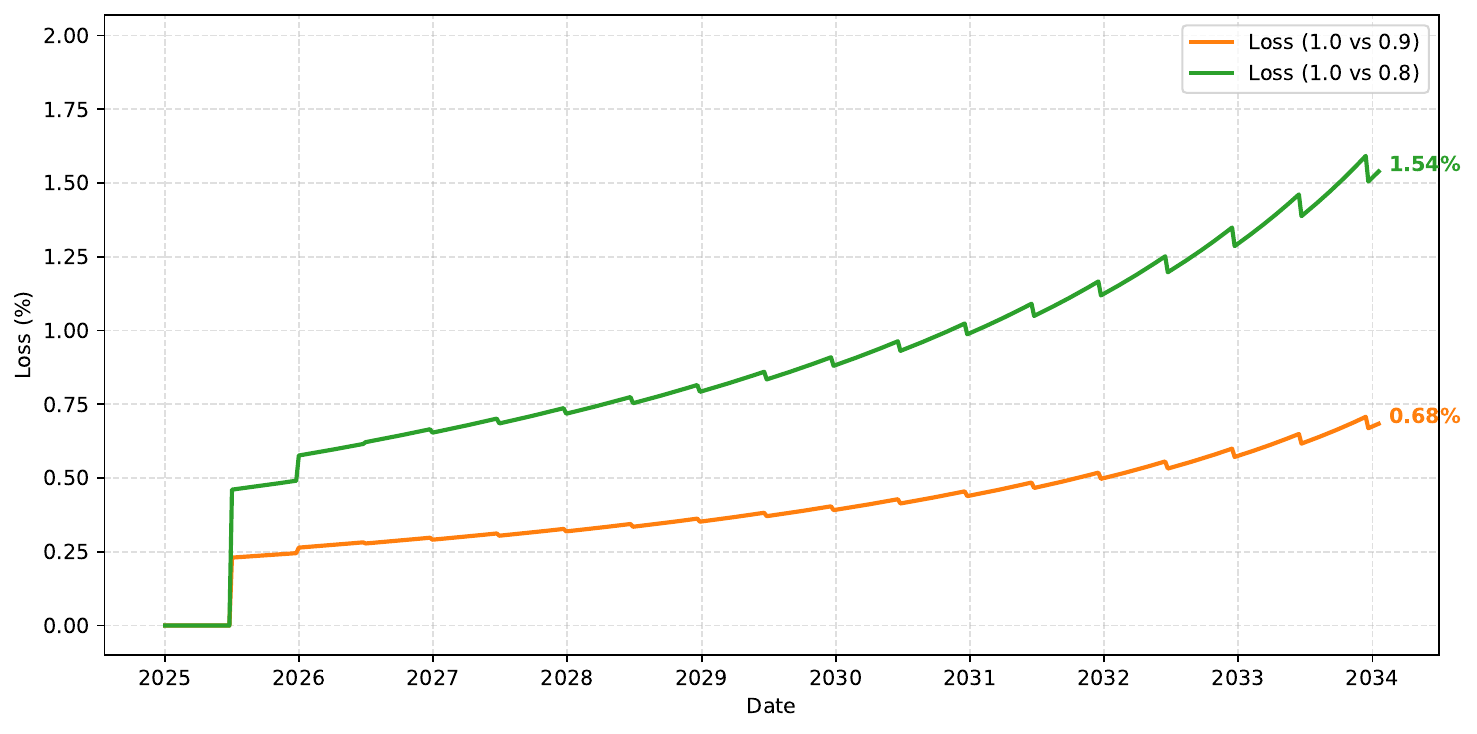}}
\caption{Cumulative financial valuation loss (\%) over time induced by information asymmetry relative to the fully accepted-record baseline}
\label{fig5}
\end{figure}

To evaluate the quantitative impact of record transparency ($\delta$), a nine-year operational scenario was simulated for a B787 aircraft assuming 3,000 annual Time in Service (TIS) hours. The restoration coefficient $\delta$ was varied from 1.0 to 0.8 to represent the spectrum from complete off-chain acceptance with blockchain-anchored provenance to fragmented conventional recordkeeping, consistent with the qualitative characterization of documentation uncertainty reported in \cite{iata2025}, \cite{iba2024}.

The simulation results indicate that the residual value differential between $\delta$ = 1.0 and $\delta$ = 0.8 reaches approximately USD 1.05 million after nine years of operation, corresponding to a 1.54\% valuation difference (Fig 4 and 5). At first glance, a 1–2\% valuation gap may appear modest. However, in the context of wide-body aircraft transactions where asset values range between USD 50–100 million in secondary markets, a USD 1 million deviation represents a financially material adjustment. Such a discrepancy directly affects lease rate factors, return condition negotiations, and portfolio-level asset valuation for lessors operating multi-aircraft fleets.

Importantly, this value differential does not arise from artificial price inflation. Rather, it reflects the elimination of verification latency between physical maintenance completion and financial recognition. In conventional systems, uncertainty regarding documentation completeness suppresses immediate value recapture, effectively reducing $\delta$ below unity during the audit window (as shown in the magnified phase of Fig. 4). By anchoring an independently accepted maintenance record at or near the time of execution, the proposed architecture may better align financial recognition with the timing of the underlying maintenance event.

Thus, transparency functions not as a speculative value enhancer, but as a friction-reduction mechanism that incrementally narrows the information asymmetry embedded within residual value assessments. Over longer operational horizons or across fleet-level portfolios, the cumulative economic effect becomes increasingly significant.

\section{Economic Impact Analysis}
To quantitatively evaluate the economic benefits of the proposed blockchain-based maintenance record management system, this section compares the audit cost structure of conventional recordkeeping systems with that of the proposed dApp. The analysis models audit cost as a function of verification workload, traceability reconstruction, operational delay, and infrastructure cost.

\subsection{Baseline Cost Model for Conventional Systems}
In traditional aircraft maintenance record audits, the total audit cost can be decomposed into four primary components:
\begin{equation}
    C^{conv}_{total} = C_{verify} + C_{trace} + C_{hangar} + C_{opportunity}
\end{equation}
where each term is defined as:
\begin{align*}
    C_{verify} = N \cdot t_{verify}\cdot r\\
    C_{trace} = N_{LLP} \cdot t_{trace} \cdot r\\
    C_{hangar} = t_{audit} \cdot c_{hangar}\\
    C_{apportunity} = t_{audit} \cdot R_{monthly}\\
\end{align*}

where N is total number of maintenance records, $t_{verify}$ is average time required to verify a single record, $r$ is labor cost per unit time, $N_{LLP}$ is number of critical records requiring full BTB traceability verification, $t_{trace}$ is average time required for record verification, $t_{audit}$ is total audit duration, $c_{hangar}$ is monthly hangar cost, and $R_{daily}$ is monthly revenue loss.

Since audit duration is fundamentally driven by record volume and traceability complexity, it can be expressed as a function of verification and traceability workload:
\begin{equation}
    t^{conv}_{audit}= \tau \cdot (N\cdot t_{verify}+ N_{LLP}\cdot t_{trace})
\end{equation}
where $\tau > 1$ represents operational delay factors.

To establish a conservative baseline, the following assumptions are adopted based on the manual, fragmented nature of conventional record validation described in \cite{iata2025}.
\begin{align*}
    t^{conv}_{verify}= 5minutes \\
    t^{conv}_{trace}= 30 minutes \\
    N_{LLP} = 0.1N
\end{align*}

\subsection{Cost Model under Blockchain-Based dApp}
The proposed system fundamentally alters the audit cost structure by introducing cryptographic verification and immutable traceability.
\begin{equation}
    C^{dApp}_{total} = C_{verify} + C_{trace} + C_{hangar} + C_{opportunity} + C_{gas}
\end{equation}
where 
\begin{equation*}
    C_{gas} = N \cdot c_{gas}
\end{equation*}
where $c_{gas}$ is the transaction cost required to anchor a single maintenance record on-chain. Based on the transaction cost analysis presented in Section III, the average blockchain transaction required to anchor a maintenance record on-chain was measured to be approximately 0.0011 BNB. To express this cost in monetary terms, the market value of BNB as of March 23, 2026, was used, corresponding to an approximate transaction cost of 0.7 USD per maintenance record.

The proposed dApp replaces human validation with cryptographic hash comparison, enabling constant-time verification independent of data size or complexity. For numerical evaluation, a conservative estimate of 0.5 minutes (30 seconds) is adopted:
\begin{equation*}
    t^{dApp}_{verify}= 0.5minutes
\end{equation*}
Unlike conventional systems, traceability reconstruction cost is negligible due to continuous and immutable linkage of maintenance records:
\begin{equation*}
    t^{dApp}_{trace} \approx 0
\end{equation*}
Thus, audit duration becomes:
\begin{equation*}
    t^{dApp}_{audit} = \tau \cdot (N\cdot t^{dApp}_{verify})
\end{equation*}

\subsection{Numerical Evaluations and Single-Aircraft Case Study}
To evaluate the practical magnitude of cost differences, a representative 10-year operational scenario is considered.

\begin{table}[htbp]
\centering
\caption{Assumptions parameters}
\label{tab:assumptions}
\renewcommand{\arraystretch}{1.6}
\begin{tabular}{cc}
\toprule
\textbf{Parameter} & \textbf{Value} \\ \midrule
$N$ & 3000 \\
$N_{LLP}$ & 300 \\
$t_{verify}^{conv}$ & 5 minutes \\
$t_{trace}^{conv}$ & 30 minutes \\
$t_{verify}^{dApp}$ & 0.5 minutes \\
$r$ & 45 USD \\
$\tau$ & 1.2 \\
$c_{hangar}$ & 5000 USD \\
$R_{monthly}$ & 256,000 USD \\
$c_{gas}$ & 0.7 USD \\ \bottomrule
\end{tabular}
\end{table}

\begin{align*}
    C^{conv}_{total} &= C_{verify} + C_{trace} + C_{hangar} + C_{opportunity}\\
    C_{verify} &= 11{,}250[USD]\\
    C_{trace} &= 6{,}750 [USD]\\
    t_{audit} &= 480hours \approx 3months\\
    C_{hangar} &= 15{,}000[USD]\\
    C_{opportunity} &= 768{,}000[USD]\\
    C^{conv}_{total} &= 801{,}000 [USD]
\end{align*}

\begin{align*}
    C^{dApp}_{total} &= C_{verify} + C_{hangar} + C_{opportunity} + C_{gas}\\
    C_{verify} &= 1{,}125[USD]\\
    t_{audit} &= 30hours \approx 1 week \approx \frac{1}{4} months\\
    C_{hangar} &= 1{,}250[USD]\\
    C_{opportunity} &= 45{,}000[USD]\\
    C_{gas} &= 2{,}250 [USD]\\
    C^{dApp}_{total} &= 49{,}625 [USD]
\end{align*}

The cost using proposed system is reduced 93\% compared to conventional maintenance system. The majority of savings arise from the drastic reduction in audit duration, which directly reduces both hangar occupancy and opportunity cost.

\subsection{Discussion }

\begin{figure}[t]
\centerline{\includegraphics[width = \columnwidth, height = 6.5cm]{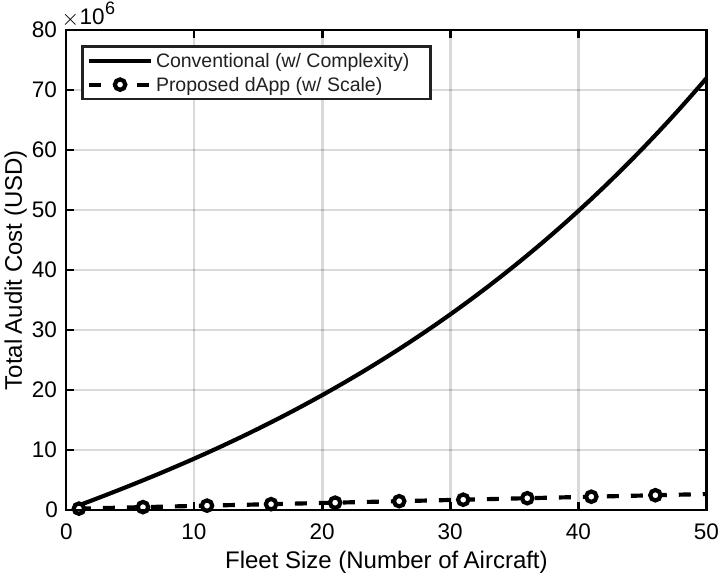}}
\caption{Total audit cost as a function of fleet size}
\label{fig6}
\end{figure}

The economic evaluation reveals that the primary driver of cost reduction is not merely labor efficiency, but the structural compression of audit duration enabled by cryptographic verification. As demonstrated in our analysis, the reduction in manual verification time per record contributes only marginally to the total savings. Instead, the elimination of traceability reconstruction—a process that typically accounts for the largest share of audit-related downtime—serves as the dominant factor.

A critical finding of this study is the divergent scaling behavior between the two systems as fleet size increases (see Fig. 6). In conventional recordkeeping, total audit costs grow super-linearly ($O(n^{1+\epsilon})$). This non-linear escalation is attributed to the accelerated complexity overhead inherent in manual cross-referencing within multi-stakeholder environments; as the number of aircraft increases, the fragmentation of maintenance data across different MROs and operators leads to a disproportionate rise in verification bottlenecks.

In contrast, the proposed dApp architecture demonstrates significant economies of scale. While it introduces a baseline infrastructure cost for system maintenance and smart contract deployment, the marginal cost per additional aircraft is drastically lower than that of conventional methods. By providing a shared, immutable provenance reference, the dApp ensures that the complexity of tracing a BtB history remains near-constant, regardless of the fleet's operational scale.

This results in an exponentially widening gap between the two curves in Fig. 6, representing cumulative cost savings that could reach several million USD for major carriers. These results indicate that while conventional auditing processes pose a significant scalability bottleneck, the proposed decentralized framework offers a predictable and optimized cost structure. Furthermore, the introduction of blockchain transaction costs ($C_{gas}$)remains negligible compared to the massive reduction in opportunity costs ($C_{opportunity}$), confirming that the computational overhead is a justified trade-off for operational agility.

Overall, the proposed system shifts the paradigm of aircraft auditing from a labor-intensive manual process to a scalable, integrity-driven automated framework. This transition is particularly vital for lessors and large-scale operators, where the cumulative impact of reduced verification latency directly translates into enhanced asset liquidity and portfolio-level economic efficiency.

\section{Conclusion}
This paper presented a blockchain-based framework for managing aircraft maintenance records with a focus on improving data integrity, traceability, and economic efficiency. By integrating off-chain storage with on-chain cryptographic anchoring, the proposed system achieves scalable data management while supporting tamper-evident verification of anchored record integrity.

A key contribution of this work lies in linking record transparency to both asset valuation and operational cost. The proposed residual value model demonstrates that maintenance record integrity can be treated as a quantifiable parameter influencing financial outcomes. In parallel, the economic impact analysis shows that the primary benefit of the system arises from audit time compression, which significantly reduces operational downtime and associated opportunity costs.

The results indicate that audit-related costs can be reduced by approximately 93\% under the stated scenario assumptions. Importantly, this reduction is not driven by labor savings alone, but by structural improvements in verification processes that eliminate traceability reconstruction and minimize delays. The benefits scale proportionally with fleet size, suggesting strong applicability for airlines operating homogeneous fleets and high-utilization aircraft.

While the proposed framework improves post-submission integrity and verification efficiency, it does not resolve the oracle problem.
False but well-formed data submitted by an authorized or compromised actor may still be immutably anchored. Therefore, the residual-value
and audit-cost benefits considered in this study apply only to records that have been accepted through appropriate off-chain technical,
organizational, or regulatory verification procedures. In the probabilistic formulation presented in Section II-C, this unresolved
source-data risk is represented by $P_{\mathrm{input}}$, which is not reduced by cryptographic hashing or blockchain consensus. Future work should therefore investigate complementary input-assurance mechanisms, including sensor-based data capture, digitally signed MRO attestations, multi-party approval, integration with trusted credential registries, and AI-assisted anomaly detection. Further empirical studies are also needed to calibrate the restoration coefficient $\delta$ using real-world lease-transition and aircraft-appraisal cases and to validate the audit-time assumptions adopted in Section IV through field measurements of record retrieval, integrity checking, semantic review, and Back-to-Birth traceability. Such empirical evidence would enable more robust estimation of both residual-value effects and operational cost savings.

Overall, this study demonstrates that blockchain-based maintenance record systems can function not only as a data integrity solution, but also as an economically impactful infrastructure that reduces inefficiencies in aircraft operations and asset management.

\section{Acknowledgement}
The authors acknowledge the use of AI-assisted tools in the preparation of this work. Gemini 3-Flash (Google)  was used to generate sample aircraft maintenance datasets for experimental validation. ChatGPT-4o (OpenAI)  was used for translation and grammatical refinement of the manuscript. Claude Sonnet 4.6 (Anthropic) was used to assist in the front-end design of the proposed decentralized application (dApp). All AI-generated outputs were reviewed, verified, and revised by the authors, who take full responsibility for the accuracy and integrity of the content presented in this paper.






\end{document}